\documentstyle[epsfig]{article}
\begin{document}

\title{Phenomenological study of solar-neutrino-induced
$\beta\beta$ process of $^{100}\mathrm{Mo}$}

\author{P. Domin$^{a)}$, F. \v Simkovic$^{a)}$,
        S.V. Semenov$^{b)}$ and Yu.V. Gaponov$^{b)}$}

\date{\today}
\maketitle
\begin{center}
{ \it 
$^{a)}$Department of Nuclear Physics, Comenius University,
SK-842 15 Bratislava, Slovakia\\
$^{b)}$Russian Research Center "Kurchatov Institute", Moscow, Russia
}
\end{center}


\begin{abstract}
The detection of solar-neutrinos of different origin
via induced $\beta\beta$ process of $^{100}\mathrm{Mo}$ is investigated.
The particular counting rates and energy 
distributions of emitted electrons are presented. 
A discussion in respect to solar-neutrino
detector consisting of 10 tones of $^{100}\mathrm{Mo}$ is
included. Both the cases of the standard solar model 
and neutrino oscillation scenarios are analyzed.
Moreover, new $\beta^-\beta^+$ and $\beta^-/EC$ channels of the
double-beta process are introduced and possibilities of their
experimental observation are addressed.
\end{abstract}

\section{Introduction}

The nature and the masses of the neutrinos has not yet been 
established phenomenologically \cite{bil99}. The problem of the 
neutrino mass scale can be solved only by combining the
neutrino oscillation data with the phenomenology of neutrinoless
double-beta decay ($0\nu\beta\beta$ decay). 

The  observed fluxes of solar neutrinos are much
reduced compared with theoretical predictions based on
the standard solar model (SSM) \cite{b&p98}. It has been confirmed
in different solar-neutrino experiments: Homestake 
\cite{HOME}, Kamiokande \cite{KAM}, 
SAGE \cite{SAGE}, GALLEX \cite{GALLEX}, and 
Super-Kamiokande \cite{SK-sun}. Recently announced result 
by the Sudbury Neutrino Observatory (SNO) has shown clear 
indication of neutrino oscillation \cite{sno}, i.e., of beyond 
standard model physics. In particular, there
is a strong evidence that $\nu_e$ from the sun are converted
to $\nu_\mu$ or $\nu_\tau$. 

The great challenge of solar-neutrino research is 
to make accurate measurements of neutrinos with energy 
less than 1 MeV. We remind that more than 98\% of the 
calculated solar-neutrino flux lies in this
energy region. Especially, the $\mathrm{pp}$ neutrinos represent 
the dominant mode of neutrino emission from the sun.
It is worthwhile to notice that the prediction of the $\mathrm{pp}$ 
neutrino flux is almost solar-model independent. New
generation of low threshold solar-neutrino experiments are in
preparation (BOREXINO) or under consideration 
(GENIUS, HERON, HELLAZ,  LENS, MOON, XMASS ...). The measurements of the total 
flux, flavor content, 
energy spectrum, and time dependence of the fundamental neutrinos 
are planned by using advantage of detection both charged and neutral weak
currents as well as of the 
neutrino-electron scattering.

In the framework of some of future experiments (GENIUS, MOON, XMASS) 
simultaneous spectroscopic studies of the $0\nu\beta\beta$ decay and 
study of low energy solar neutrinos  will  be performed. In  the MOON
detector with 10 tons of $^{100}\mathrm{Mo}$  solar neutrinos are proposed
to be detected by inverse $\beta$ decay \cite{eji00}. We note that
$^{100}\mathrm{Mo}$ is suitable for charge current registration of
solar neutrinos due to  low threshold of $0.168~\mathrm{MeV}$,  
what allows observation of low energy  $\mathrm{pp}$ neutrinos. 

In this contribution we analyze the possibility of detection of
solar neutrinos via neutrino-induced double-beta process 
($\beta\beta$-process) of $^{100}\mathrm{Mo}$. 
We note that the $\beta\beta$ process induced 
by neutrinos was discussed for the first time in connection with detection of 
reactor neutrinos and artificial sources in Refs. \cite{sem98,inzh98}. 
Here, we present the 
predictions for the solar neutrinos absorption rates by assuming the
standard solar model and different neutrino oscillation scenarios.
The subject of our interest is also the background coming from the 
two-neutrino double-beta decay of $^{100}\mathrm{Mo}$ in registration of 
solar neutrinos of different origin. Finally, new modes of induced 
$\beta\beta$ process, $\beta^-\beta^+$ and $\beta^-/EC$ channels, 
are introduced  and discussed.

\section{Cross-section of the neutrino induced $\beta\beta$ process}

The neutrino induced double-beta transition (induced $\beta\beta$ transition) 
\cite{sem98},
\begin{equation}
\nu_e + (A,Z) \rightarrow (A,Z+2) + 2 e^- + {\overline{\nu}}_e,
\label{eq:1}
\end{equation}
is a reaction allowed within the Standard model. It is a second order
process in the weak interaction, observation of which by the help of 
reactor-neutrino flux has been found possible due to resonant enhancement 
of the corresponding amplitude depending on the width of the 
intermediate nuclear state \cite{inzh98}. Below, we briefly present the
theory of the induced $\beta\beta$ process concerning ground state to ground 
state nuclear transition. 

The differential cross section of the induced $\beta\beta$ process 
can written as
\begin{equation}
\frac{d \sigma(\varepsilon_{\nu})}{d \varepsilon_1 d \varepsilon_2} =
\frac{(G_{\beta})^{4} g_A^4 m_e^4}{8 \pi^5} 
\pi_1 \varepsilon_1 F(\varepsilon_1, Z_f) 
\pi_2 \varepsilon_2 F(\varepsilon_2, Z_f) 
(w_0 - \varepsilon_1 - \varepsilon_2 + \varepsilon_{\nu})^2 K_{GT}
\label{eq:cs}
\end{equation}
with $\pi_k=p_k/m_e$, $\varepsilon_k=E_k/m_e$ ($k=1$, $2$)
and $\varepsilon_{\nu}=E_\nu/m_e$. $p_k$ and $E_k$ 
are momenta and energies of electrons, respectively. $E_{\nu}$ denotes 
energy of the incident neutrino.  $F(\varepsilon_1, Z_f)$ 
is Coulomb correction factor, 
$w_0$ is the energy difference of the target and final
nuclei in units of electron mass.  Factor $K_{GT}$ containing the
energy denominators is given by 
\begin{equation}
K_{GT} = \frac{3}{4} \left( \left|\sum\limits_{m} (K_m+L_m)\right|^2+
\frac{1}{3} \left| \sum\limits_{m} (K_m - L_m) \right|^2 \right)
\label{eq:KGT}
\end{equation}
with
\begin{eqnarray}
K_m &=& \left[(\varepsilon_m + \frac{1}{2}i\gamma_m - \varepsilon_i 
+ \varepsilon_1 - \varepsilon_{\nu})^{-1} + 
(\varepsilon_m + \frac{1}{2}i\gamma_m - \varepsilon_i + \varepsilon_2 
+ \varepsilon_{\overline{\nu}_2})^{-1}\right] \nonumber \\
& \times & \langle 0^+_{g.s.}, f || \sum\limits_{i} \tau_i^+ \vec{\sigma}_i || 1^+_m \rangle 
\langle 1^+_m || \sum\limits_{i} \tau_i^+ \vec{\sigma}_i || 0^+_{g.s.}, i 
\rangle, 
\nonumber \\
L_m &=& \left[(\varepsilon_m + \frac{1}{2}i\gamma_m - \varepsilon_i + 
\varepsilon_1 + \varepsilon_{\overline{\nu}_2})^{-1} + 
(\varepsilon_m + \frac{1}{2}i\gamma_m - \varepsilon_i + \varepsilon_2 
- \varepsilon_{\nu})^{-1}\right] \nonumber \\
& \times & \langle 0^+_{g.s.}, f || \sum\limits_{i} \tau_i^+ \vec{\sigma}_i || 1^+_m \rangle 
\langle 1^+_m || \sum\limits_{i} \tau_i^+ \vec{\sigma}_i || 0^+_{g.s.}, i \rangle. 
\label{eq:denom}
\end{eqnarray}
Here, $\varepsilon_m = E_m/m_e$ and $\gamma_m = \Gamma_m/m_e$. 
$E_m$ and $\Gamma_m$ are  energy and width of the $m$-th $1^+$ 
intermediate nuclear state, respectively. $\varepsilon_{\overline{\nu}_2}$
is the energy of outcoming entineutrino in unit of $m_e$. 
$| 0^+_{g.s.}, i \rangle$, $| 0^+_{g.s.}, f \rangle$ 
and $ |1^+_m \rangle$ denote wave functions of initial, final and intermediate 
nuclear states, respectively.

The factors $K_m$ and $L_m$ are similar to those entering the expression
for two-neutrino double-beta decay ($2\nu\beta\beta$ decay)
amplitude with one important  difference. 
The sign in front of the energy of incoming neutrino is changed to minus.
Than two of four denominators in Eq.  (\ref{eq:denom}) exhibit
singular behavior if the incident neutrino energy is 
larger than the threshold for creation of a real state of the intermediate 
nucleus 
\begin{equation}
\varepsilon_{\nu} > 1+\varepsilon_m-\varepsilon_i, \quad  (m < m^{RS}).
\label{eq:cond}
\end{equation}
$\varepsilon_{m^{RS}}$ denotes the energy of highest lying excited state of the 
intermediate nucleus satisfying relation (\ref{eq:cond}).
 
The total cross-section is calculated by keeping in mind that widths of low 
lying states of the intermediate nucleus are very small in comparision with
their energies due to electromagnetic and weak decay channels. 
After performing integration over the energies of outgoing electrons 
in Eq. (\ref{eq:cs}), we obtain
\begin{eqnarray}
\sigma(\varepsilon_{\nu}) &=& 
\frac{(G_{\beta})^4 g_A^4 m_e^6}{2 \pi^4} 
\sum\limits_{m=0}^{m^{RS}}
\left| \langle 0^+_{g.s.}, f || \sum\limits_{i} \tau_i^+ \vec{\sigma}_i 
|| 1^+_m \rangle
\langle 1^+_m || \sum\limits_{i} \tau_i^+ \vec{\sigma}_i 
|| 0^+_{g.s.}, i \rangle \right|^2 \nonumber \\
&\times&
\frac{1}{\gamma_m} 
\pi_r^{(m)} \varepsilon_r^{(m)} 
F(Z_f,\varepsilon_r^{(m)})
\int\limits_{1}^{\varepsilon_m-\varepsilon_f} d\varepsilon_2
\pi_2 \varepsilon_2 F(Z_f,\varepsilon_2)
(\varepsilon_m-\varepsilon_f-\varepsilon_2)^2,
\end{eqnarray}
where $\varepsilon_r^{(m)}=\varepsilon_{\nu}-\varepsilon_m+\varepsilon_i$. 
We note that the dominant contribution to the 
cross-section comes from the transitions through the real states of the
intermediate nucleus ($m < m^{RS}$) and that  contribution of
transitions through
 virtual intermediate states (like in $2\nu\beta\beta$ decay)
is negligible (suppressed by factor $\sim \gamma_m$). 


\begin{figure}[!ht]
\centerline{\epsfig{file=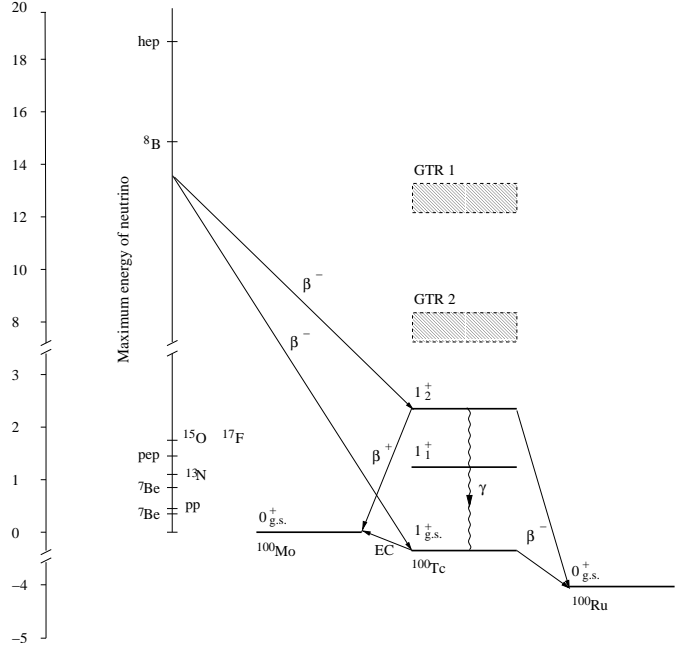,width=0.7\linewidth}}
\caption{Nuclear level and transition schemes $^{100}\mathrm{Mo}$ 
for $\beta\beta$ processes induced by solar neutrino absorption. 
GTR1 and GTR2  denote giant Gamow-Teller resonances. 
The energy scale is in units of MeV.
\label{fig:1}}
\end{figure}

Next, we introduce the partial width $\gamma_{mf}$ of $m$-th state 
of the intermediate nucleus associated with the $\beta^-$ transition to 
the ground state of final nucleus. We have
\begin{eqnarray}
\gamma_{mf} = 
\frac{(G_{\beta})^2 g_A^2 m_e^4}{2\pi^3}
\left| \langle 0^+_{g.s.}, f || \sum\limits_{i} \tau_i^+ \vec{\sigma}_i 
|| 1^+_m \rangle \right|^2 
\int\limits_{1}^{\varepsilon_{m}-\varepsilon_f}
d\varepsilon_2
\pi_2 \varepsilon_2 F(Z_f,\varepsilon_2)
(\varepsilon_{m_0}-\varepsilon_f-\varepsilon_2)^2 \label{eq:gmf}.
\end{eqnarray}
It allows us to write the total cross-section of the neutrino-induced
$\beta\beta$-process in a compact form
\begin{equation}
\sigma(\varepsilon_{\nu}) = \sum\limits_{m=0}^{m^{RS}}
\sigma_{\beta}^{(m)}(\varepsilon_{\nu}) \frac{\gamma_{mf}}{\gamma_{m}},
\label{eq:cs2}
\end{equation}
where $\sigma_{\beta}^{(m)}(\varepsilon_{\nu})$ represents 
the cross-section of the neutrino-induced single-$\beta$ decay from the ground 
state of the initial nucleus to the $m$-th state of the intermediate nucleus: 
\begin{equation}
\sigma_{\beta}^{(m)}(\varepsilon_{\nu})=
\frac{(G_{\beta})^2 g_A^2 m_e^2}{\pi}
\left| \langle 1^+_m || \sum\limits_{i} \tau_i^+ \vec{\sigma}_i
|| 0^+_{g.s.}, i \rangle \right|^2
\pi^{(m)}_r \varepsilon^{(m)}_r F(Z_f,\varepsilon^{(m)}_r). 
\label{eq:sb}
\end{equation}

The total width $\gamma_m$ of the $m$-th excited state of the intermediate 
nucleus is much larger as the partial $\beta$ width $\gamma_{mf}$ due
to preferably electromagnetic deexcitation to 
lower lying states. However, in the case of ground state of the 
intermediate nucleus there is a different situation, in particular
${\gamma_{mf}}/{\gamma_{m}}\approx 1$. 
Thus the transition through the lowest state of the intermediate nucleus 
gives the dominant contribution to the total cross section of the
induced $\beta\beta$ process (see Eq. (\ref{eq:cs2})). We note that a
different situation can be in case of the neutrino-induced $\beta\beta$
process with emission of $\gamma$-ray, which has origin in electromagnetic
deexcitation of levels of the intermediate nucleus. The details of this
process we expect to discuss in next publication.

\begin{table}[t]
\caption{Solar neutrino signal rates per day
in $10$ tones of 
$^{100}\mathrm{Mo}$ for transition through ground state as well as 
excited states of intermediate $^{100}\mathrm{Tc}$. 
B(GT) and $E^*$ denote the Gamow-Teller strength and
the excitation energy, respectively. 
\label{tab:1}}
\begin{center}
\begin{tabular}{|l|lllll|}
\hline
state:& $1^+_{g.s.}$&$1^+_{1}$&$1^+_{2}$&GTR1&GTR2\\
$B(GT)$:&$0.33$&$0.13$&$0.23$&$23.1$&$2.9$\\
$E^{*} (MeV)$:&$0$&$1.4$&$2.6$&$13.3$&$8.0$\\
\hline
source& 
\multicolumn{5}{c|}
{Production rate in $10$ tones of $^{100}\mathrm{Mo}$ per day}\\
\hline
pp &$3.4$&-&-&-&-\\
pep &$7.1~10^{-2}$&-&-&-&-\\
$^{7}{\mathrm Be}(1)$&$0.04$&-&-&-&-\\
$^{7}{\mathrm Be}(2)$&$1.0$&-&-&-&-\\
$^{8}{\mathrm B}$&$0.04$&$0.01$&$0.01$&$1.4~10^{-4}$&$0.01$\\
hep&$3.1~10^{-5}$&$9.6~10^{-6}$&$1.3~10^{-5}$& $2.1~10^{-5}$&$3.8~10^{-5}$\\
$^{13}{\mathrm N}$&$0.11$&-&-&-&-\\
$^{15}{\mathrm O}$&$0.16$&$5.1~10^{-4}$&-&-&-\\
$^{17}{\mathrm F}$&$1.96~10^{-3}$&$6.4~10^{-6}$&-&-&-\\
\hline
\end{tabular}
\end{center}
\end{table} 

From the above study it follows that electrons emitted
in the neutrino-induced $\beta\beta$ process exhibit different
behavior from those  in the $2\nu\beta\beta$ decay, which 
is considered to be the major background in some of planned 
solar-neutrino detectors. The induced $\beta\beta$ transition 
can be to a good
approximation  considered as two subsequent independent
processes (see Eq. (\ref{eq:cs2})) the induced single-$\beta^-$ transition 
to the ground state of
intermediate nucleus followed by nuclear $\beta^-$ decay to the ground
state of final nucleus. The time delay between both $\beta$
processes is about 20 second for A=100 system. It is worthwhile to notice
that in the $2\nu\beta\beta$ decay electrons are emitted 
simultaneously. In addition, they are strongly correlated what
allows to eliminate $2\nu\beta\beta$ decay background 
through coincidence measurements.

\section{Solar neutrino absorption rates for 
$^{100}\mathrm{Mo}$}

The solar neutrino absorption rates for $^{100}\mathrm{Mo}$ are
important in the evaluation of detector for planned MOON experiment.
In this section we present complementary and more detailed 
calculations to those of Ref. \cite{eji00}.

The absorption rate of solar neutrinos from the source $s$ is given by 
\cite{b&p98,jnb}
\begin{equation}
R_{\nu} = \int \sigma(\varepsilon_{\nu}) \rho_{s}(\varepsilon_{\nu}) 
d \varepsilon_{\nu}.
\end{equation}
Here, $\rho_{s}(\varepsilon_{\nu})$ is the spectrum of incident 
solar-neutrino flux. The total cross-section $\sigma(\varepsilon_{\nu})$ 
of the neutrino-induced 
$\beta\beta$ process is given in Eq. (\ref{eq:cs2}).

\begin{figure}[t]
\centerline{\epsfig{file=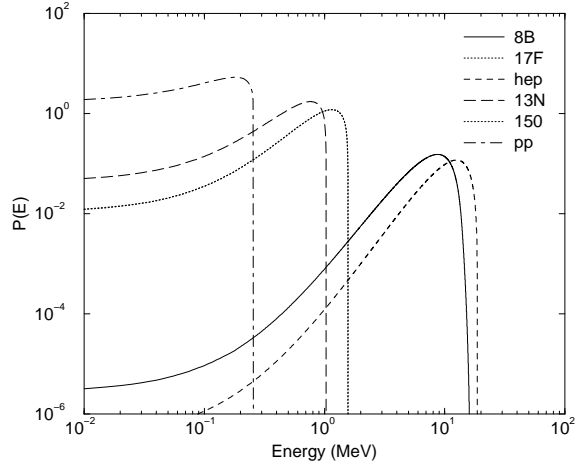,width=0.6\linewidth}}
\caption{Probability density of the emission of the single electron via  
solar-neutrino-induced inverse $\beta\beta$ decay
on $^{100}\mathrm{Mo}$. \label{fig:2}}
\end{figure}

The nuclear level scheme for solar-neutrino-induced
$\beta\beta$-transition of $^{100}\mathrm{Mo}$  is presented in  Fig. \ref{fig:1}.
The dominant nuclear transitions are realized through  $1^+$ states
of the intermediate nucleus. The corresponding Gamow-Teller strengths 
have been determined experimentally by measuring the electron
capture and $\beta$-decay rates of $^{100}\mathrm{Tc}$ ($\log ft_{\beta}=4.60$)
\cite{sin98}  as well as of cross-section of the 
$^{100}\mathrm{Mo}(^{3}\mathrm{He},t)^{100}\mathrm{Tc}$ 
reaction \cite{acu97}. 

\begin{table}[t]
\caption{Particular solar-neutrino 
capture rates 
in unit of SNU (solar-neutrino unit) for 
$^{100}\mathrm{Mo}$, $^{37}\mathrm{Cl}$ and $^{71}\mathrm{Ga}$.
Contributions coming only from  the dominant transition through the 
ground state of the intermediate nucleus
$^{100}\mathrm{Tc}$ were taken into account. 
\label{tab:2} 
}
\begin{center}
\begin{tabular}{|lllllllll|}\hline
source & $pp$ & $pep$ & $^{7}\mathrm{Be}$ & $^{8}\mathrm{B}$ & $hep$ &
$^{13}\mathrm{N}$ & $^{15}\mathrm{O}$ &  $^{17}\mathrm{F}$ \\ \hline
$^{100}\mathrm{Mo}$  & 652  & 13.6 & 197  & 7.8  & 0.01 & 22   & 31.6 & 0.38 \\
$^{37}\mathrm{Cl} $  & 0    & 0.22 & 1.15 & 5.76 & 0.04 & 0.09 & 0.33 & 0.0  \\
$^{71}\mathrm{Ga} $  & 69.7 & 2.8  & 34.2 & 12.1 & 0.1  & 3.4  & 5.5  & 0.1  \\
\hline
\end{tabular}
\end{center}
\end{table} 

In Table (\ref{tab:1}) we present  solar-neutrino
signal rates per day in $10$ tones of 
$^{100}\mathrm{Mo}$, assuming there are no oscillations 
of neutrinos. We note that transitions through higher excited
states of the intermediate nucleus are strongly suppressed.
There transitions were estimated by assuming $\gamma$-deexcitation to the
ground state of the intermediate nucleus.
We remind that neutrino-induced $\beta\beta$ transition through excited 
intermediate states (without accompanying $\gamma$-ray emission) 
is strongly suppressed due 
to small ratio ${\gamma_{mf}}/{\gamma_{m}}$. 

In  calculation we used energy distribution of  solar neutrinos
predicted by the standard solar model \cite{b&p98,jnb}.
The probability density 
for emission of single electron of given energy  
induced by solar neutrinos of
a given source is plotted in Fig. (\ref{fig:2}).
The solar-neutrino  absorption rate for $^{100}\mathrm{Mo}$ is
compared with those for $^{37}\mathrm{Cl}$ and $^{71}\mathrm{Ga}$
isotopes
(see Table  \ref{tab:2}).   We note  
that detector built of $^{100}\mathrm{Mo}$ isotope allows 
to register significantly larger amount of  solar neutrinos
as it is in the case of $^{37}\mathrm{Cl}$ and $^{71}\mathrm{Ga}$. It is 
due to lower threshold for solar-neutrino
absorption by this isotope. The relevant thresholds 
for $^{100}\mathrm{Mo}$,  $^{37}\mathrm{Cl}$ 
and $^{71}\mathrm{Ga}$ targets are 
$0.168~ MeV$, $0.817~MeV$ and $0.235~MeV$, respectively.

It is worthwhile to notice
 that there are also other standard model allowed channels
of the neutrino-induced $\beta\beta$-transition 
($\beta^-/EC$ and $\beta^-\beta^+$ modes), which have been not
discussed in literature yet. For 
$^{100}\mathrm{Mo}$ target they are given by
\begin{eqnarray}
\nu_e + ^{100}\mathrm{Mo} \to ^{100}\mathrm{Mo} + e^- + e^+ + \nu_e
~~(\beta^-/EC), \nonumber \\
\nu_e + e^- + ^{100}\mathrm{Mo} \to ^{100}\mathrm{Mo} + e^- + \nu_e
~~(\beta^-\beta^+). 
\end{eqnarray} 
The corresponding cross-sections of these reactions can be calculated by 
omitting the term $L_m$ in Eqs. (\ref{eq:cs}), (\ref{eq:KGT}) and
(\ref{eq:denom}). In addition,
in the case of $\beta^-/EC$ mode one has to replace $\varepsilon_2$ 
with $-1$ and to modify appropriate the normalization of one electron. 
The cross-sections of these modes are of the form given in 
Eq. (\ref{eq:cs2}), however, $\gamma_{mf}$ represents the partial width 
of  $m$-th intermediate $1^+$ state of $^{100}\mathrm{Tc}$
in respect to $\beta^+$ or $EC$ decay channels. For the ground state
of  $^{100}\mathrm{Tc}$ we have: 
$\gamma_{mf}/\gamma_m~=~ 0.0018~ (EC), ~0~ (\beta^+)$. Thus, 
$\beta^- /EC$ mode is suppressed in comparison with the
$\beta^-\beta^-$ one by about of three orders of magnitude.
In the case of $\mathrm{pp}$ solar neutrinos 
the corresponding   absorption rate is $1.2$~SNU 
($2.3$ events per year in $10$~tones of $^{100}\mathrm{Mo}$). 
The $\beta^-\beta^+$ mode is undetectable for this type of
detector.

Nowadays, we have a strong indication in favor of solar neutrino
oscillations coming from SNO experiment \cite{sno}. There are different
scenarios for mixing of neutrinos, which differ very little
in their predictions. Usually,  one vacuum oscillation solution 
and three MSW solutions of the 
deficit of solar neutrino events are considered. They are 
the Small Mixing Angle, the Large Mixing Angle, the LOW $\Delta m^2$ solutions. 
In Table \ref{tab:3} 
we discuss the solar neutrino deficit
assuming SMA, LMA, LOW and vacuum oscillation schemes  for neutrinos
of different origin. We notice that LMA and LOW solar neutrino
absorption rates are close each to other for all types of neutrinos.
The SMA solution offers considerably smaller values especially
in the case of $\mathrm{pp}$, $\mathrm{pep}$ and $^{7}\mathrm{Be}$ neutrinos. 

\begin{table}[t]
\caption{The solar neutrino absorption rates in 
$^{100}\mathrm{Mo}$ in SNU. The standard solar model (SSM) predictions
are compared with those of three MSW and one vacuum oscillation solutions 
of the solar neutrino problem~\cite{bah98}. SMA, LMA, LOW and VO stand for 
the small mixing angle, the large mixing angle,
the low $\Delta m^2$ and vacuum oscillations solutions,
respectively.
\label{tab:3} }
\begin{center}
\begin{tabular}{|lllllllll|}
\hline
source&
pep&
$^{7}\mathrm{Be}(1)$&
$^{7}\mathrm{Be}(2)$&
pp&
$^{8}\mathrm{B}$&
$^{13}\mathrm{N}$&
$^{15}\mathrm{O}$&
$^{17}\mathrm{F}$\\
\hline
SSM&
$14$&
$0.81$&
$197$&
$652$&
$7.8$&
$22$&
$32$&
$0.38$\\
&&&&&&&&\\
SMA&
$0.05$&
$0.41$&
$0.58$&
$540$&
$2.9$&
$1.09$&
$0.50$&
$0.01$\\
LMA&
$4.97$&
$0.45$&
$90$&
$372$&
$2.19$&
$10$&
$13$&
$0.15$\\
LOW&
$6.00$&
$0.44$&
$94$&
$351$&
$3.18$&
$11$&
$14$&
$0.17$\\
VO &
$0.5$&
$0.55$&
$162$&
$100$&
$0.63$&
$3$&
$6$&
$0.07$\\
\hline
\end{tabular}
\end{center}
\end{table}

\section{Conclusions}
 
The solar-neutrino anomaly is a  controversial problem,  
which attracts attention of both theoreticians and experimentalists for
a long time. New experimental data and theoretical ingredients
have to be merge to give reliable predictions for oscillation
parameters.
 
In this contribution the charged current
solar neutrino absorption  is discussed in 
context of the neutrino-induced $\beta\beta$ process. 
We have shown that to a good approximation, 
this process can be considered as neutrino-induced 
$\beta$-decay followed by  single-$\beta$ decay. This fact
allows to identify clearly the 
$2\nu\beta\beta$ decay background
in planned solar neutrino experiments.

Moreover, complementary theoretical study in respect to the MOON
project is presented. Solar-neutrino absorption rates
are calculated by assuming SSM neutrino flux spectrum. 
The probability density for emission of single electron
is evaluated. A comparison of the SSM prediction 
for absorption of neutrinos in $^{100}\mathrm{Mo}$ is compared 
with those of most favored neutrino oscillation scenarios.

\end{document}